\documentclass[aip,amsmath,amssymb,preprint]{revtex4-1}

\usepackage{graphicx}
\usepackage{dcolumn}
\usepackage{bm}

\usepackage[utf8]{inputenc}
\usepackage[T1]{fontenc}
\usepackage{mathptmx}
\usepackage{etoolbox}

\renewcommand{\AA}{${\mathring{A}}$}
\renewcommand{\r}{\vec{r}}

\newcommand{\ket}[1]{\left|#1\right>}
\newcommand{\bra}[1]{\left<#1\right|}
\newcommand{\bk}[2]{\left<#1|#2\right>}
\newcommand{\bak}[3]{\left<#1|#2|#3\right>}
\newcommand{\erf}[1]{{\textrm{ erf}}\left(#1\right)}
\newcommand{\erfc}[1]{{\textrm{ erfc}}\left(#1\right)}

\makeatletter
\def\@email#1#2{%
 \endgroup
 \patchcmd{\titleblock@produce}
  {\frontmatter@RRAPformat}
  {\frontmatter@RRAPformat{\produce@RRAP{*#1\href{mailto:#2}{#2}}}\frontmatter@RRAPformat}
  {}{}
}%
\makeatother
\begin{document}

\preprint{AIP/123-QED}

\title[Projected Hybrids ]{Projected Hybrid Density
Functionals: Method and Application to Core Electron Ionization} 
\author{Benjamin G. Janesko}
 \email{b.janesko@tcu.edu}

\affiliation{ Department of Chemistry \& Biochemistry, Texas Christian University, Fort Worth, TX 76129, USA}

\date{\today}

\begin{abstract} This work presents a new class of hybrid density functional
theory (DFT) approximations, incorporating nonlocal exact exchange in predefined
states such as core atomic orbitals (AOs). 
These projected hybrid density functionals are a flexible generalization of
range-separated hybrids.
This work derives projected hybrids using the Adiabatic Projection formalism.
One projects the electron-electron interaction operator onto the chosen
predefined states, reintroduces the projected operator into the noninteracting
Kohn-Sham reference system, and introduces a density functional approximation
for the remaining electron-electron interactions.
Projected hybrids are readily implemented existing density functional codes,
requiring only a projection of the one-electron density matrices and exchange
operators entering existing routines. 
This work also presents a first application: a core-projected
Perdew-Burke-Ernzerhof hybrid PBE0c70, in which the fraction of nonlocal exact
exchange is increased from 25\% to 70\% in core AOs. Automatic selection of
the projected AOs provides a black-box model chemistry appropriate for both
core and valence electron properties.  PBE0c70 predicts core orbital energies
that accurately recover core-electron binding energies of second- and third-row
elements,
without degrading PBE0's good performance for valence-electron properties. 
\end{abstract} 

\maketitle 

\section{Introduction}

Kohn-Sham density functional theory (DFT) is the most widely-used electronic
structure approximation across chemistry, physics, and materials science.\cite{Verma2020} DFT
models a system of interacting electrons in terms of a reference system of
noninteracting Fermions, corrected by a mean-field (Hartree) electron repulsion
and an exchange-correlation (XC) density functional incorporating all many-body
effects.\cite{Janesko2021}
Standard approxmations to the XC functional can capture important aspects of
covalent bonding, at the expense of delocalization and self-interaction errors
that lead to overbinding and
over-delocalization.\cite{MoriSanchez2008,Perdew2017}
Hybrid XC approximations mitigate these effects by including a fraction of
nonlocal exact exchange, effectively reintroducing part of the
electron-electron interaction into the reference system.\cite{Toulouse2009}
However, introducing a {\em{fixed}} fraction of the {\em{entire}}
electron-electron interaction leads to pervasive and resilient zero-sum
tradeoffs between underbinding and
over-delocalization.\cite{Ruzsinszky2006,Janesko2017,Janesko2021}


The generalized range-separated adiabatic connection provides a way to optimize
these
tradeoffs.\cite{Savin1988,Savin1995,Leininger1997,Yang1998,Toulouse2004,Fromager2007,Toulouse2009,Savin2020,Pernal2021}
This approach separates the electron-electron interaction operator $\hat{V}_{ee}$ into
short-range and long-range pieces, for example 
\begin{eqnarray}
\label{eq:range}
\hat{V}_{ee} &=& \sum_{i>j}\frac{1}{r_{ij}} \ = \ \sum_{i>j} \frac{\erf{\mu r_{ij}}}{r_{ij}} + \frac{\erfc{\mu r_{ij}}}{r_{ij}}, \\ 
&=& \hat{V}^{LR}_{ee} + \hat{V}^{SR}_{ee}.  \nonumber 
\end{eqnarray}
Part of the interaction, typically the long-range part, is reintroduced into
the noninteracting reference system. 
The Hohenberg-Kohn theorems ensure that the real system's ground-state energy
and density can be obtained from an exact wavefunction calculation on the
long-range-interacting reference system, corrected by a density functional for
the remaining short-range Hartree-exchange-correlation energy.
By relying on approximate XC functionals for only {\em{part}} of the
electron-electron interaction, these approaches can provide beyond-zero-sum
accuracy for certain properties, without the expense of a correlated
wavefunction calculation on the real system.\cite{Pernal2021}
Several groups have explored different approximations for the long-range-interacting reference system
wavefunction, including coupled-cluster theory,\cite{Goll2005} multireference
approaches,\cite{Fromager2007,Fromager2010} the density matrix renormalization
group,\cite{Hedegaard2015} and the random phase
approximation.\cite{Janesko2009b}
 
Range-separated hybrids are an especially widely adopted
approach.\cite{Henderson2008a} These methods approximate the reference system
wavefunction as a single Slater determinant, corrected by a density functional
for full-range correlation.\cite{Toulouse2009}
The development of range-separated exchange
functionals\cite{Iikura2001,Ernzerhof1998} has enabled broad adoption of
range-separated hybrids. Long-range-corrected (LC) hybrids introduce the
long-range interaction into the reference system, and are widely adopted for
Rydberg and charge-transfer excited states, noncovalent interactions, and
more.\cite{Dreuw2003,Hedegaard2013,Rohrdanz2009}  Screeneed hybrids introduce
the short-range interaction into the reference system, and are widely adopted
for semiconductors, metal oxides, and core excitations.\cite{Heyd2003,Wang2020}
However, introducing a {\em{fixed}} range separation $\mu$ and {\em{fixed}}
fraction of the short- and long-range interactions leaves remaining
zero-sum tradeoffs.\cite{Janesko2017} Recent efforts to treat these tradeoffs
include system-dependent range separation,\cite{Stein2009,Koerzdoerfer2014}
local range separation,\cite{Krukau2008} range separations parameterized to
particular properties,\cite{Besley2009} and other approximations reviewed in
ref \citenum{Janesko2021}. 

\subsection{Adiabatic Projection}

The Adiabatic Projection approach generalizes the
range-separated adiabatic connection.\cite{Janesko2022} One replaces the
range-separate interaction  of eq \ref{eq:range} with a projected interaction
defined by a set of two-electron projection operators $\{\hat{P}^{(2)}_m\}$: 
\begin{eqnarray}
\label{eq:Vp}
\hat{V}^P_{ee} &=& \sum_m c_m \sum_{i>j}\hat{P}^{(2)}_m(i,j) \frac{1}{r_{ij}} \hat{P}^{(2)}_m(i,j). 
\end{eqnarray}
(The notation $\hat{P}_m^{(2)}(i,j)$ means that the operator acts on electrons $i$ and $j$.) 
One then reintroduces the projected interaction into the reference system.  
Just as for the range-separated adaibatic connection, the
Hohenberg-Kohn theorems ensure that the real system's exact ground-state energy
and density can be obtained from an exact wavefunction calculation on the
projected-interacting reference system, corrected by a formally exact density
functional for the projected Hartree-exchange-correlation energy.

We have applied the Adiabatic Projection approach to several contemporary
problems in DFT. 
Projecting onto one-electron states
$\hat{P}^{(2)}_m=\ket{\phi_m\phi_m}\bra{\phi_m\phi_m}$,
$\bk{\r}{\phi_m}=\phi_m(\r)$ introduces {\em{only self-interaction}}
into the reference system. Choosing those one-electron states as localized
Kohn-Sham spin-orbitals, and approximating the projected XC functional in terms of
the orbital densities, recovers the Perdew-Zunger self-interaction
correction (PZSIC).\cite{Perdew1981,Janesko2022a}  Other choices of one-electron states
provides connections between the PZSIC, the Hubbard model DFT+U, and Rung 3.5
approximations.\cite{Janesko2022b} Projecting onto active spaces of multiple
occupied and virtual orbitals, and treating the reference system with a complete
active-space self-consistent field (CASSCF) wavefunction, generalizes the PZSIC
into a wavefunction-in-DFT approach.\cite{Janesko2022a}

\subsection{Projected hybrids} 

This work introduces projected hybrid
density functionals inspired by range-separated hybrids. One
projects the electron-electron interaction onto predefined states such as core
atomic orbitals (AOs), approximates the reference system wavefunction as a
single Slater determinant, and combines a projected exchange
functional\cite{Janesko2022a} with a full-range correlation functional. This
flexible approach permits exact exchange admixture in chemically appropriate
regions, requires minimal modification to existing codes, and (unlike
active-space approaches) can be incorporated into ``black-box'' model
chemistries.

\subsection{DFT for core electron properties}
\label{sec:core} 

This pilot study introduces a core-projected hybrid functional targeted to
simulate core- and valence- electron properties. 
Core-electron spectroscopies probe the chemical environment of nuclei and give
element-specific information on bonding and oxidation state. 
The growing availability of bright X-ray sources has led to increasing interest
in core electron spectroscopies.\cite{Norman2018}
DFT and time-dependent (TD-)DFT simulations are widely adopted
to interpret core electron spectra.\cite{Besley2020}
DFT orbital energies, generated from accurate Kohn-Sham or generalized Kohn-Sham\cite{Seidl1996}
potentials, can accurately predict the vertical ionization potentials (IPs) of both
core and valence electrons.\cite{Chong2002,Bellafont2015,Verma2012a} (In this approach, the first
ionization potential is modeled as the negative of the highest occupied
molecular orbital energy IP=-$\epsilon_{HOMO}$, and core ionization potentials are
modeled as the negative of core molecular orbital energies.) Accurate XC
potentials and orbital energies are also important for linear response TD-DFT
simulations of X-ray absorption spectra.\cite{Zhang2012,Verma2016} 
 
Self-interaction error significantly impacts DFT-predicted core electron
properties. For second-row elements Li-Ne, the core orbital energies predicted
by standard global hybrid functionals differ by tens of eV from
experimental K-edge core IPs.\cite{Tu2007} Comparable errors occur for TD-DFT
predictions of core excitation energies.\cite{Nakata2006,Maier2016} 
Self-interaction errors are even worse for the more compact cores of third-row
elements Na-Ar.\cite{Besley2021} 
Increasing the admixture of nonlocal exact exchange can improve predicted core
IPs at the expense of a ``zero-sum'' degradation in valence electron
properties.\cite{Fouda2020} 
Delta-self-consistent-field ($\Delta$SCF) approaches explicitly compute
non-Aufbau core-ionized or core-excited state wavefunctions,\cite{Besley2009a}
providing a reduced impact of self-interaction error and a widely adopted
practical solution.\cite{Norman2018}  However, $\Delta$SCF approaches can suffer
from difficulties converging the non-Aufbau states, require one SCF calcuation
for each nucleus of interest in a large molecule, do not readily yield
transition moments or vibronic couplings, and may still be significantly
impacted by self-interaction error in heavier elements.\cite{Besley2021} 
The state-of-the-art for modeling X-ray fluorescence combines linear response
TD-DFT with a constant shift taken from $\Delta$SCF
calculations.\cite{Fouda2020} Mitigating the impact of self-interaction error
on core elections, without degrading the treatment of valence electrons, could
broaden the impact of inexpensive TD-DFT approaches.\cite{Maier2016} 

There has been significant interest in using self-interaction correction or
exact exchange admixture to improve TD-DFT predictions of core electron
spectroscopies.
Tu and coworkers predicted core ionization potentials by applying a rescaled
PZSIC to B3LYP-computed orbital energies.\cite{Tu2007}  
Several workers have developed range-separated hybrids that incorporate a large
fraction of short-range nonlocal exchange. 
Hirao and coworkers modified their LCgau-BOP hybrid to include additional
short-range nonlocal exchange, and reported improved TD-DFT core excitations for
second-row atoms.\cite{Song2007,Song2008} 
Besley and coworkers reparameterized screened hybrid functionals to improve
TD-DFT predictions of core excitations. They required different
parameterizations for second-row and third-row atoms,\cite{Besley2009} which may be 
another manifestation of the zero-sum tradeoffs discussed above. 
Chai and coworkers introduced short- and long-range correction (SLC) hybrids
including 100\% nonlocal exchange at short and long range. SLC core orbital
energies accurately predicted core ionization energies of second-row atoms, and
TD-DFT with SLC functionals accurately predicted a range of core, valence, and
charge-transfer excitations.\cite{Wang2016} 
Kaupp and coworkers showed that local hybrid functionals, incorporating a
position-dependent admixture of exact exchange, provided balanced accuracy for
TD-DFT predictions of the core excitations of second-row elements,  along with
valence, charge-transfer and Rydberg excitations.\cite{Maier2016} 
Nakata and coworkers introduced an orbital-dependent hybrid incorporating
different fractions of HF exchange in different Kohn-Sham orbitals. The authors
reported TD-DFT calcuations using the coupling operator technique of Roothaan,
and found accurate core-excitation energies for second-row
atoms.\cite{Nakata2006}  This work was extended to a core-valence-Rydberg
approach,\cite{Nakata2006a} and is related to other orbital-dependent DFT
methods.\cite{Ranasinghe2015}

\subsection{Core projected hybrids} 

This pilot study presents projected hybrids that incorporate additional nonlocal 
exchange in core AOs. Enhancing the Perdew-Burke-Ernzherhof global hybrid
PBE0\cite{Perdew1996,Adamo1999,Scuseria1999} with 70\% exact exchange in core
AOs yields core molecular orbital (MO) energies that accurately predict second-
and third-row $K$-edge ionization potentials, without a zero-sum degradation in
valence electron properties.  Numerical results are comparable to the
core-valence hybrid of Nakata and coworkers,\cite{Nakata2006} without requiring
the cumbersome coupling operator technique. The rest of this work presents a
derivation of projected hybrids, details of the core-projected hybrid tested
here, and numerical results. 

\section{Derivation} 

This derivation of projected hybrid density functionals is 
based on published derivations of range-separated hybrids.\cite{Toulouse2009}
In Kohn-Sham DFT, the exact ground-state energy of an $N$-electron system is
expressed in terms of a noninteracting reference system and a density
functional correction: 
\begin{eqnarray}
\label{eq:ref}
E &=& \min_\Phi \left(\bak{\Phi}{\hat{T}+\hat{V}_{ext}}{\Phi} + E_{HXC}[\rho_\Phi]\right). 
\end{eqnarray}
Here $\hat{T}$ and $\hat{V}_{ext}$ are operators for the the kinetic energy and
external potential experienced by the reference system. $E_{HXC}[\rho]$ is the
universal Hartree-exchange-correlation (HXC) density functional. Given the
exact HXC functional, the reference system's minimizing single-determinant
wavefunction $\Phi$ yields ground-state energy $E$ and electron density $\rho_\Phi$ equal to
the exact values. 

Range-separated and projected hybrids respectively introduce the
range-separated (eq \ref{eq:range}) and projected (eq \ref{eq:Vp}) interactions
into the reference system. The present work introduces a new choice of the
projection in eq \ref{eq:Vp}, based on an orthonormal set of $N_{P}$
single-particle states $\{\phi^{(P)}_m\}$ that are independent of the Kohn-Sham
orbitals: 
\begin{eqnarray}
\label{eq:Pcore}
\hat{P}^{(2)}&=&\sum_{m,n}^{N_{P}}
\ket{\phi^{(P)}_m\phi^{(P)}_n}\bra{\phi^{(P)}_m\phi^{(P)}_n}. 
\end{eqnarray}
(For example, core-projected hybrids will choose  $\{\phi^{(P)}_m\}$ by
orthogonalizing the core AOs.) 
The ground-state energy is expressed as 
\begin{eqnarray}
\label{eq:LR}
E &=& \min_{\Psi^{LR}} \left(\bak{\Psi^{LR}}{\hat{T}+\hat{V}_{ext} +\hat{V}_{ee}^{LR}}{\Psi^{LR}} + E_{HXC}^{SR}[\rho_\Psi^{LR}]\right), 
\end{eqnarray}
in the generalized range-separated adiabatic connection and as 
\begin{eqnarray}
\label{eq:P0}
E &=& \min_{\Psi^{P}} \left(\bak{\Psi^{P}}{\hat{T}+\hat{V}_{ext} +\hat{V}_{ee}^{P}}{\Psi^{P}} + E_{HXC}^{P}[\rho_\Psi^{P}]\right) , 
\end{eqnarray}
in the present work. 
Minimizing wavefunctions $\Psi^{LR}$ and $\Psi^{P}$ are generally
multideterminant. Eq \ref{eq:LR}-\ref{eq:P0} yield the exact density and energy
of the real system given the exact short-range and projected HXC functionals,
respectively. The short-range HXC functional depends on the chosen range
separation ($\mu$ in eq \ref{eq:range}), and the projected HXC functional
depends on the projection $\hat{P}^{(2)}$ and thus on the chosen
$\{\phi^{(P)}_n\}$. 

Range-separated and projected hybrids are derived by restricting the minimizing
wavefunctions in eq \ref{eq:LR}-\ref{eq:P0} to be single determinant: 
\begin{eqnarray}
\label{eq:RSH0}
E^{(0)}_{RSH} &=& \min_{\Phi^{LR}} \left(\bak{\Phi^{LR}}{\hat{T}+\hat{V}_{ext} +\hat{V}_{ee}^{LR}}{\Phi^{LR}} + E_{HXC}^{SR}[\rho_\Phi^{LR}]\right), \\ 
\label{eq:PH0}
E^{(0)}_{PH} &=& \min_{\Phi^{P}} \left(\bak{\Phi^{P}}{\hat{T}+\hat{V}_{ext} +\hat{V}_{ee}^{P}}{\Phi^{P}} + E_{HXC}^{P}[\rho_\Phi^{P}]\right). 
\end{eqnarray}
The minimizing determinants are given by the Euler-Lagrange equations: 
\begin{eqnarray}
\label{eq:ellr}
\left(\hat{T}+\hat{V}_{ext}+\hat{J}-(\hat{K}-\hat{K}^{SR})+\hat{V}_{XC}^{SR}\right)\ket{\Phi^{LR}} &=& \epsilon_{LR}\ket{\Phi^{LR}}, \\ 
\label{eq:elp}
\left(\hat{T}+\hat{V}_{ext}+\hat{J}-(\hat{K}-\hat{K}^{P})+\hat{V}_{XC}^{P}\right)\ket{\Phi^{P}} &=& \epsilon_{P}\ket{\Phi^{P}}.
\end{eqnarray}
Here $\hat{J}$ is the Hartree potential, the sum of long-range and short-range
terms (eq \ref{eq:ellr}) or projected and unprojected terms (eq \ref{eq:elp}).
$\hat{K},\hat{K}_{SR},\hat{K}_{P}$ are the full-range, short-range, and
projected nonlocal exchange operators. $\epsilon_{LR}$ and $\epsilon_{P}$ are
Lagrange multipliers for the normalization constraint. 
 
The restriction to single determinants means that eq \ref{eq:RSH0}-\ref{eq:PH0} do not 
yield the exact energy and density, even with the exact short-range or
projected Hartree-exchange-correlation functionals. Nevertheless, they can be
used as a reference to express the exact energy as $E=E_{RSH}+E_C^{LR}$ or
$E_{PH}+E_C^{P'}$. Conventional LC hybrids use a standard (full-range)
correlation functional to model the sum of correlation in the
long-range-interacting reference system $E_C^{LR}$, and the
correlation piece of $E_{HXC}^{SR}$.\cite{Toulouse2009} This work uses a standard
(unprojected) correlation functional to model the sum of correlation in the
projected-interacting reference system, and the correlation piece of 
$E_{HXC}^P$. The hybrid functionals' the total energies are 
\begin{eqnarray}
E_{RSH} &=& \min_{\Phi^{LR}} \left(\bak{\Phi^{LR}}{\hat{T}+\hat{V}_{ext} +\hat{V}_{ee}^{LR}}{\Phi^{LR}} + U_{SR}[\rho_\Phi^{LR}]+ E_{XSR}^{SL}[\rho_\Phi^{LR}] + E_{C}[\rho_\Phi^{LR}]\right), 
\end{eqnarray}
for range-separated hybrids and 
\begin{eqnarray}
\label{eq:PH}
E^{(0)}_{PH} &=& \min_{\Phi^{P}} \left(\bak{\Phi^{P}}{\hat{T}+\hat{V}_{ext} +\hat{V}_{ee}^{P}}{\Phi^{P}} + U_P[\rho_\Phi^{P}] + E_{XP}^{SL}[\rho_\Phi^{P}] + E_C^{SL}[\rho_\Phi^{P}]\right), 
\end{eqnarray}
for projected hybrids. 

Practical implementation of range-separated hybrid functionals requires 
explicit construction of range-separated semilocal exchange functionals
$E_{XSR}^{SL}[\rho]$, based on models for the exchange hole.
This is not the case for projected exchange functionals, where one may pass
projected one-particle density matrices to unmodified semilocal exchange
functionals. Ref \citenum{Janesko2022a} provides a detailed derivation of this
approach, as a generalization of the Perdew-Zunger self-interaction correction.

\section{Implementation} 

This section presents the working equations for a projected hybrid's
generalized Kohn-Sham energy and potential. 
We consider the case of a core-projected hybrid calculation in an atomic
orbital (AO) basis. 
The reference system's single-determinant $N$-electron wavefunction
$\ket{\Phi^P}$ is made up of orthonormal one-electron spin-orbitals (MOs)
$\{\ket{\psi_{i\sigma}}\}$, each of which is
expanded in a nonorthogonal basis set of $N_{basis}$ real AOs $\{\ket{\chi_\mu}\}$. 
The overlap $\bk{\chi_\mu}{\chi_{\nu}}=S_{\mu\nu}$, where 
$S_{\mu\nu}$ is a matrix element of the
$N_{basis}\times N_{basis}$ overlap matrix ${\bf{S}}$.  
The MOs are expanded as
$\ket{\psi_{i\sigma}}=\sum_\mu
c_{i\sigma\mu}\ket{\chi_\mu}$. 
The core-projected hybrid has $N_{core}\ll N_{basis}$
AOs are assigned as core AOs. The orthonormal projection states in eq
\ref{eq:Pcore} are denoted core states $\{\ket{\phi^{(c)}_m}\}$, and  are obtained by
orthogonalizing the core AOs. Orthogonalization requires
$\left(S^{(c)}\right)^{-1}_{\mu\nu}$, the matrix inverse of the $N_{core}\times
N_{core}$ matrix of core AO overlaps. 

Proceeding requires a one-electron projection
operator $\hat{P}$ onto the orthogonalized core AOs. 
This operator's
representation in the full AO basis is
\begin{eqnarray}
\label{eq:P}
\hat{P} &=& \sum_{\mu\nu}\ket{\chi_\mu}P_{\mu\nu}\bra{\chi_{\nu}}. 
\end{eqnarray}
Matrix element $P_{\mu\nu}$ of $N_{basis}\times N_{basis}$ matrix ${\bf{P}}$
equals $\left(S^{(c)}\right)^{-1}_{\mu\nu}$ when both $\chi_\mu$ and $\chi_\nu$
are core AOs, and equals zero elsewhere. This matrix obeys
${\bf{PSP}}={\bf{P}}$.  The two-electron operator in eq \ref{eq:Vp} becomes
$\hat{P}^{(2)}(i,j)=\hat{P}(i)\hat{P}(j)$, where  $\hat{P}(i)$ denotes the 
projection of eq \ref{eq:P} acting on electron $i$. 
For any one-electron operator $\sum_i\hat{A}(i)$, the expectation value of the
projected operator $\sum_i\hat{P}(i)\hat{A}(i)\hat{P}(i)$ becomes
\begin{eqnarray}
\bak{\Phi}{\sum_i\hat{P}(i)\hat{A}(i)\hat{P}(i)}{\Phi} &=& \sum_{i\sigma}\bra{\psi_{i\sigma}}
\left(\sum_{\mu\mu'}\ket{\chi_{\mu}}P_{\mu\mu'}\bra{\chi_{\mu'}}\right)\hat{A}
\left(\sum_{\nu'\nu}\ket{\chi_{\nu'}}P_{\nu'\nu}\bra{\chi_{\nu}}\right) 
\ket{\psi_{i\sigma}},\\ 
&=& \sum_{i\sigma}\sum_{\mu\nu}c_{i\sigma\mu}^{*}A^P_{\mu\nu} c_{i\sigma\nu} . \nonumber 
\end{eqnarray}
This equation introduces matrix elements of the projected one-electron
operator, which are defined as 
\begin{eqnarray}
\label{eq:Ap}
A^P_{\mu\nu} &=& \sum_{\mu'\mu"\nu'\nu"} S_{\mu\mu'}P_{\mu'\mu"}A_{\mu"\nu"}P_{\nu"\nu'}S_{\nu'\nu}. 
\end{eqnarray} 
$A_{\mu\nu}=\bak{\chi_\mu}{\hat{A}}{\chi_\nu}$ denotes matrix elements of
the unprojected operator. The expectation value of the projected two-electron
integral operator $\hat{V}_{ee}^P=\sum_{i\geq j}
\hat{P}(i)\hat{P}(j)\hat{V}_{ee}(i,j)\hat{P}(i)\hat{P}(j)$ 
 is 
\begin{eqnarray}
\label{eq:VeeP}
\bak{\Phi}{\hat{V}_{ee}^P}{\Phi} &=& \frac{1}{2}\sum_{i\sigma} \sum_{j\sigma'}\bra{\psi_{i\sigma}\psi_{j\sigma'}}
\left(\sum_{\mu\nu\mu'\nu'}\ket{\chi_{\mu}\chi_{\nu}}P_{\mu\mu'}P_{\nu\nu'}\bra{\chi_{\mu'}\chi_{\nu'}}\right)\hat{V}_{ee}\\ 
&\times&
\left(\sum_{\lambda'\eta'\lambda\eta}\ket{\chi_{\lambda'}\chi_{\eta'}}P_{\lambda'\lambda'}P_{\eta'\eta}\bra{\chi_{\lambda}\chi_{\eta}}\right) 
\left(\ket{\psi_{i\sigma}\psi_{j\sigma'}}-\ket{\psi_{j\sigma'}\psi_{i\sigma}} \right) ,\nonumber \\ 
&=& \frac{1}{2}\sum_{i\sigma}\sum_{j\sigma} c_{i\sigma\mu}^*c^*_{j\sigma'\nu}\left<\mu\nu|\lambda\eta\right>^P 
\left(c_{i\sigma\lambda}c_{j\sigma'\eta}-c_{j\sigma'\lambda}c_{i\sigma\eta}\right) .\nonumber 
\end{eqnarray}
The projected two-electron integrals are defined analogous to eq \ref{eq:Ap}: 
\begin{eqnarray}
\left<\mu\nu|\lambda\eta\right>^P &=& \sum_{\mu'\mu"\nu'\nu"}\sum_{\lambda'\lambda"\eta'\eta"} 
S_{\mu\mu'}P_{\mu'\mu"}
S_{\nu\nu'}P_{\nu'\nu"}
\left<\mu"\nu"|\lambda"\eta"\right>
P_{\lambda"\lambda'}S_{\lambda'\lambda}
P_{\eta"\eta'}S_{\eta'\eta}
\end{eqnarray}
In practice, implementing projected hybrids does not require explicitly
evaluating these projected AO-basis two-electron integrals.  One only requires
one-electron projections of density matrices and exchange operators. (However,
the projected AO-basis two-electron integrals may be required for
multi-determinant treatments of the projected interacting reference system,
discussed in sec \ref{sec:discussion}.)
As shown in eq \ref{eq:elp}, the Hartree piece of eq \ref{eq:VeeP}
is combined with the remainder of the Hartree interaction from the
projected Hartree-exchange-correlation density functional, yielding the usual
${\bf{J}}$ matrix. 
%
The exchange piece of eq \ref{eq:VeeP} can be written as 
\begin{eqnarray}
\label{eq:ExP}
E_{X}^P &=& -\frac{1}{2}\sum_{ij\sigma} \sum_{\mu\nu\lambda\eta} c^*_{i\sigma\mu}c^*_{j\sigma\nu}\left<\mu\nu|\lambda\eta\right>^P c_{j\sigma\lambda} c_{i\sigma\eta}, \\ 
&=& -\frac{1}{2}\sum_{i\sigma} \sum_{\mu\eta} c^*_{i\sigma\mu} K^P_{\mu\eta}[\gamma^P_\sigma]c_{i\sigma\eta} .\nonumber  
\end{eqnarray}
Here $K^P_{\mu\eta}[\gamma^P_\sigma]$ denotes the projected (as in eq
\ref{eq:Ap}) nonlocal exact exchange operator constructed from projected
density matrix $\gamma^P_\sigma$. Matrix elements of the the conventional AO-basis $\sigma$-spin
one-particle density matrix are
\begin{eqnarray}
\gamma_{\sigma\nu\lambda} &=& \sum_i c_{i\sigma\nu}c^*_{i\sigma\lambda}. 
\end{eqnarray}
The projected density matrix is defined as 
\begin{eqnarray}
\label{eq:gammaP}
\gamma_{\sigma\nu\lambda}^P &=&  \sum_{\nu'\nu"\lambda'\lambda"} P_{\nu\nu'}S_{\nu'\nu"}\gamma_{\sigma\nu"\lambda"}S_{\lambda"\lambda'}P_{\lambda'\lambda}. 
\end{eqnarray} 
(Note that the projected density matrix eq \ref{eq:gammaP} is ${\bf{P S \gamma
S P}}$, whereas the projected one-electron operator eq \ref{eq:Ap} is ${\bf{S P
A P S }}$.) 
The unprojected exchange operator is constructed from the projected density
matrix and unprojected AO-basis two-electron integrals as, 
\begin{eqnarray}
K_{\mu\eta}[\gamma^P] &=& \sum_{\nu"\lambda"}\left<\mu\nu"|\lambda"\eta\right>\gamma^P_{\sigma\nu"\lambda"} 
%
%
\end{eqnarray}
Projecting this operator as in eq \ref{eq:Ap} recovers the first line of eq
\ref{eq:ExP}.

The final step in the implementation involves the projected exchange
functional, the analogue of the range-separated exchange functionals used in
range-separated hybrids.  As suggested above, we obtain the projected exchange
functional by passing the projected density matrix to existing semilocal exchange functionals.
We consider a standard $\sigma$-spin semilocal exchange functional 
\begin{eqnarray}
\label{eq:EXSL} 
E_{X}^{SL}[\gamma_\sigma] &=& \int d^3\r \ e_{X}^{SL}\left[\rho_\sigma(\r),|\nabla\rho_\sigma(\r)|\ldots\right], \\ 
\rho_\sigma(\r) &=& \sum_{\mu\nu}\chi_\mu(\r)\gamma_{\mu\nu\sigma}\chi_\nu(\r). 
\end{eqnarray}
We define the projected exchange functional as the difference between
$E_X^{SL}$ evaluated with projected vs. unprojected density matrices. 
The total energy of a projected hybrid functional thus becomes 
\begin{eqnarray}
\label{eq:E}
E &=& \sum_{ij\sigma} \sum_{\mu\nu}c_{i\sigma}^*\left(h_{\mu\nu}+\frac{1}{2}J_{\mu\nu}-\frac{1}{2}K^P_{\mu\nu}[\gamma^P_\sigma]\right) + E_{XC}^{SL}[\gamma] -\sum_\sigma E_X^{SL}[\gamma^P_\sigma]
\end{eqnarray}
Here $h_{\mu\nu}$ is a matrix element of the kinetic and external potential
operators, $J_{\mu\nu}$ is a matrix element of the standard full-range Hartree
potential, $E_{XC}^{SL}[\gamma]$ is a standard semilocal exchange-correlation
functional evaluated on the full density matrix, and
$E_X^{SL}[\gamma^P_\sigma]$ is the exchange piece of the semilocal functional
(eq \ref{eq:EXSL}) evaluated on the projected density matrix of eq
\ref{eq:gammaP}. The Fock-like matrix defined by $\partial E/\partial
c_{i\mu\sigma} =\sum_\nu F_{\mu\nu\sigma}c_{i\nu\sigma}$, becomes  (compare
with eq \ref{eq:elp}) 
\begin{eqnarray}
\label{eq:F}
F_{\mu\nu\sigma} &=& h_{\mu\nu} + J_{\mu\nu} + \left(V_{XC}^{SL}[\gamma]\right)_{\mu\nu\sigma} -K^P_{\mu\nu}[\gamma^P_\sigma]-\left(V_X^{SLP}[\gamma^P_\sigma]\right)_{\mu\nu}
\end{eqnarray}
Here the unprojected semilocal exchange operator is constructed from the
projected density matrix in the usual way 
\begin{eqnarray}
\left(V_X^{SL}[\gamma^P_\sigma]\right)_{\mu\nu} &=& \int d^3\r \chi_\mu(\r) \frac{\delta e_{X\sigma}^{SL}}{\delta \rho^P_\sigma(\r)} \chi_\nu(\r) + \ldots 
\end{eqnarray}
Projecting this operator as in eq \ref{eq:Ap} recovers the
$\left(V_X^{SLP}[\gamma^P_\sigma]\right)_{\mu\nu}$ in eq \ref{eq:F}. 
 Operationally, one constructs  the projected AO-basis density matrix ${\bf{P S
\gamma S P}}$, passes this to standard routines for constructing ${\bf{K}}$ and
the semilocal exchange potential, then projects the operators (e.g. ${\bf{S P K
P S}}$) before use.  Because the projection states are independent of the MOs,
self-consistent implementation merely requires these projections of
one-particle density matrices and exchange operators.

\section{Methods} 

This work uses an implementation of eq \ref{eq:E}-\ref{eq:F} into the PySCF\cite{Sun2020}
electronic structure package. The implementation is freely available online at
github.com/bjanesko. Just as range-separated hybrids can include different
fractions of short- and long-range exact exchange, the present implementation
includes different fractions of projected exact exchange $\alpha_c$ and global
exact exchange $\alpha_0$.  This work adopts the notation "FcX", where "F" is a
standard XC functional and "X" is the fraction of nonlocal exchange in core
AOs. X="HF" is equivalent to X=100. For example, PBEcHF combines semilocal PBE
with 100\% HF exchange in AOs.  Most benchmark calculations treat the PBE0c70
functional, combining PBE0 with 70\% nonlocal exchange in core AOs
($\alpha_0=25\%,\alpha_c=70\%$). 

In this pilot study, total energies and Fock-like matrices (eq \ref{eq:F}) are
computed from Hartree-Fock one-particle density matrices. Generalized Kohn-Sham
orbital energies are obtained from a single diagonalization of the Fock-like
matrices constructed from Hartree-Fock density matrices. 
Most calculations use the Perdew-Burke-Ernzerhof\cite{Perdew1996} (PBE)
generalized gradient approximation, the PBE0 global
hybrid incoporating 25\% nonlocal exchange,\cite{Adamo1999,Scuseria1999} the
PBEHH global hybrid incorporatng 50\% nonlocal exchange, or the HFPBE
combination of nonlocal exchange and PBE correlation. Test calculations treat
the Becke-Lee-Yang-Parr (BLYP) GGA,\cite{Lee1988,Becke1988} the three-parameter
global hybrid B3LYP,\cite{Becke1993a,Cheeseman1996} the
Tao-Perdew-Staroverov-Scuseria (TPSS) and Strongly Constrained and
Appropriately Normed (SCAN) meta-GGAs,\cite{Tao2003,Sun2015} and the SCAN0
global hybrid. 
Other test calculations use the Gaussian 16 package,\cite{g16} and also include the
M06L, M06, M06-2X, and M06-HF global
hybrids,\cite{Zhao2006,Zhao2006a,Zhao2006c} the HSE06, N12SX, and MN12XS screened
hybrids,\cite{Heyd2003,Peverati2012c} and the LC-$\omega$PBE, $\omega$B97X-D,
M11, Lc-BLYP, and CAM-B3LYP long-range-corrected
hybrids.\cite{Vydrov2006a,Yanai2004,Peverati2011} These calculations use
post-HF total energies and self-consistent orbital eneriges.
For the systems tested here, the two approaches methods are nearly identical:
single-shot PBE with PySCF predicts CH$_3$Cl valence and core IP 7.14 and
2738.9 eV, self-consistent PBE with Gaussian 16 predicts valence and core IP
7.09 and 2739.7 eV. 
Calculations use several AO basis sets, including the 6-311++G(2d,2p)
Pople-type basis set,\cite{Ditchfield1971,Hehre1972} the cc-pVnZ correlation-consistent basis
sets,\cite{Dunning1989a} the cc-pCVTZ and cc-pCVQZ basis sets designed for core electron
properties, and the def2-SVP, def2-TZVP, def2-QZVP basis
sets.\cite{Weigend2005}
All molecular geometries are B3LYP/6-311++G(2d,2p) optimized. 

This black-box study includes an automated choice of core AOs.  Each second-
and third-row element is assigned a cutoff kinetic energy as 1.4 times the
kinetic energy from the most diffuse uncontracted core orbital in the STO-2G
basis set. All s-type contrated AOs with kinetic energy above that cutoff are
assigned to the core. For example, the STO-2G basis set for lithium atom
includes a contracted 1s AO with exponents 6.16 and 1.10 au. The most diffuse
exponent gives a kinetic energy $(3/2)1.10$ au and a cutoff $(1.4)(3/2)1.10$
au=2.3 au. All s-type contracted AOs centered on a Li atom $\{\chi_\mu^{Li}\}$,
whose kinetic energy expectation value
$\bak{\chi_\mu^{Li}}{\frac{1}{2}\nabla^2}{\chi_\mu^{Li}}$ is above that
threshold, are assigned as core. For the 6-311++G(2d,2p) basis set, this
approach gives one core AOs for each second-row element and three core AOs for
each third-row element. 

\section{Results} 

\subsection{Validation} 

Table \ref{tab:basis} illustrates the overall performance of this approach, as
well as the basis set dependence. The table shows valence, Cl core, and C core
ionization potentials of CH$_3$Cl, computed from the corresponding generalized
Kohn-Sham orbital energies. Calculations compare the PBEcHF core-projected
hybrid (100\% nonlocal exchange in core AOs) with PBE (no nonlocal exchange) and
HFPBE (100\% nonlocal exchange globally). As expected, the core-projected hybrid recovers
the PBE valence IP to within 0.01 eV, and recovers the HFPBE core IP to within
a few percent. Results are robust across basis sets, even as the number of core
AOs changes from 2 to 11.  This confirms that the projected hybrids and the
core AO selection process perform as expected.

\renewcommand{\arraystretch}{0.8}
\begin{table}
\caption{\label{tab:basis} Valence, C core, and Cl core ionization potentials
of methylene chloride CH$_3$Cl (eV), computed as the negative of GKS orbital
energies, evaluated in various basis sets.}
\begin{tabular*}{\textwidth}{@{\extracolsep{\fill}}l r rrr rrr rrr} 
\hline\hline
 & & \multicolumn{3}{c}{Valence }& \multicolumn{3}{c}{C atom core } & \multicolumn{3}{c}{Cl atom core } \\ 
Basis & $N_{core}$ & PBE & PBEcHF & HFPBE  & PBE & PBEcHF & HFPBE  & PBE & PBEcHF & HFPBE  \\ 
\hline 
STO-3G & 2 & 4.94 & 4.94 & 11.64 & 267.3 & 303.5 & 303.7 & 2709.0 & 2814.1 & 2822.1\\
3-21G & 2 & 7.09 & 7.09 & 13.15 & 270.5 & 307.2 & 306.9 & 2719.0 & 2826.7 & 2833.2\\
6-31G(d) & 2 & 7.04 & 7.04 & 13.09 & 271.8 & 307.8 & 308.2 & 2738.4 & 2844.3 & 2852.7\\
6-311++G(2d,2p) & 4 & 7.14 & 7.14 & 13.17 & 272.0 & 311.3 & 308.4 & 2738.9 & 2847.7 & 2853.2\\
cc-pvdz & 4 & 6.95 & 6.95 & 13.03 & 272.0 & 307.7 & 308.4 & 2738.7 & 2847.6 & 2853.0\\
cc-pvtz & 3 & 7.05 & 7.05 & 13.09 & 271.8 & 308.5 & 308.2 & 2738.8 & 2847.6 & 2853.1\\
cc-pcvtz & 5 & 7.06 & 7.06 & 13.09 & 271.8 & 307.8 & 308.2 & 2738.8 & 2847.9 & 2853.1\\
cc-pvqz & 5 & 7.09 & 7.09 & 13.11 & 271.8 & 308.5 & 308.2 & 2738.9 & 2847.8 & 2853.2\\
cc-pcvqz & 11 & 7.09 & 7.09 & 13.11 & 271.8 & 307.8 & 308.2 & 2738.9 & 2847.3 & 2853.2\\
def2-svp & 2 & 6.79 & 6.79 & 12.93 & 272.0 & 308.0 & 308.3 & 2737.5 & 2843.8 & 2851.8\\
def2-tzvp & 3 & 7.06 & 7.06 & 13.08 & 271.8 & 299.9 & 308.2 & 2738.8 & 2848.2 & 2853.1\\
def2-qzvp & 6 & 7.10 & 7.10 & 13.12 & 271.8 & 303.7 & 308.2 & 2738.9 & 2848.5 & 2853.2\\
\hline\hline
\end{tabular*}
\end{table}

Table \ref{tab:meth} illustrates core-projected hybrid calculations using a
variety of standard XC functionals. All core-projected hybrids use 100\%
nonlocal exchange in cores.  For HFPBE, the unprojected and core-projected
functionals are identical by construction. Otherwise, core projection increases
the predicted core IP of second- and third-row atoms, without much affecting
the HOMO energies.  Whereas the unprojected functionals' predicted core IP
range over 30 eV for C and 110 eV for Cl, the core-projected functionals' core
IP are all within 6 eV of each other.  

\begin{table} \caption{\label{tab:meth} Valence, C core, and Cl core ionization
potentials of methylene chloride (eV), computed as the negative of GKS orbital
energies, unmodified XC functional F vs. core-projected functional FcHF,
6-311++G(2d,2p)
basis set.}
\begin{tabular*}{\textwidth}{@{\extracolsep{\fill}}l rr rr rr}
\hline\hline 
  & \multicolumn{2}{c}{Valence }& \multicolumn{2}{c}{C atom core } & \multicolumn{2}{c}{Cl atom core } \\ 
Functional F & F & FcHF & F & FcHF & F& FcHF \\
\hline
PBE & 7.14 & 7.14 & 272.0 & 311.3 & 2738.9 & 2847.7\\
PBE0 & 8.64 & 8.64 & 281.1 & 310.6 & 2767.4 & 2849.0\\
PBEHH & 10.15 & 10.15 & 290.2 & 309.8 & 2796.0 & 2850.4\\
HFPBE & 13.17 & 13.17 & 308.4 & 308.4 & 2853.2 & 2853.2\\
BLYP & 6.96 & 6.96 & 272.7 & 311.6 & 2740.7 & 2848.6\\
B3LYP & 8.17 & 8.17 & 279.7 & 311.3 & 2762.4 & 2849.9\\
TPSS & 7.33 & 7.33 & 274.8 & 310.7 & 2747.9 & 2848.6\\
SCAN & 7.55 & 7.55 & 275.9 & 310.3 & 2753.6 & 2850.3\\
SCAN0 & 8.89 & 8.89 & 284.0 & 309.7 & 2778.4 & 2851.0\\
\hline\hline
\end{tabular*} \end{table}

\subsection{Benchmarks}

\begin{figure}[ht]
\includegraphics[width=0.8\textwidth]{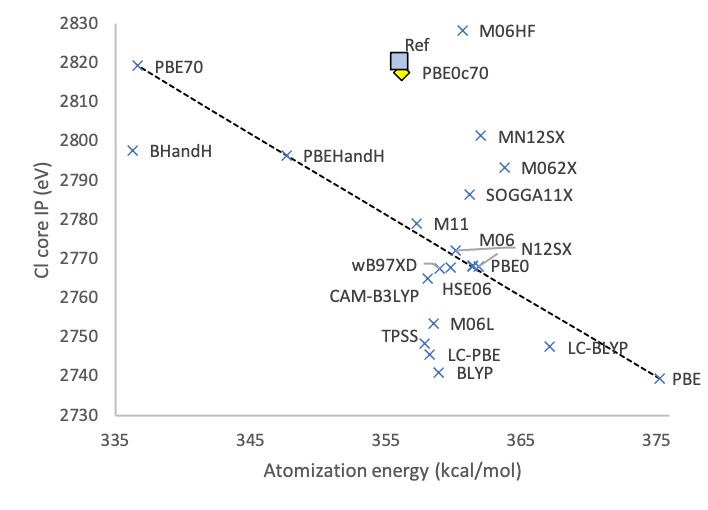}
\caption{\label{fig:zero} Computed atomization energy (abscissa) and Cl core IP
(ordinate) of formyl chloride HCOCl. "X" denotes standard DFT functionals. Straight line
denotes global hybrids of PBE including between 0 and 70\% exact exchange.
"Ref" is accurate reference values. }
\end{figure}

Figure \ref{fig:zero} highlights the ``zero-sum'' trade-offs between
predictions of valence and core properties, focusing on the atomization energy
and core IP of formyl chloride HCOCl. Calculations use the def2-TZVP basis
set. The figure shows the atomization energy on the abscissa, and the Cl core
IP (negative of computed Cl 1s orbital energy) on the ordinate.  Reference
values are the 2820.6 eV core IP from ref \citenum{Besley2021}, and a new
CBS-QB3\cite{Montgomery1999} computed atomization energy 355.9 kcal/mol.
Calculations compare core-projected PBE0c70 to a broad range of density
functionals: semilocal PBE, BLYP, TPSS, M06L; global hybrid PBE0, PBEHandH,
BHandH, M06-2X, M06-HF, and SOGGA11-X; and range-separated hybrids M11,
$\omega$B97X-D, HSE06, LC-$\omega$PBE, LC-BLYP, CAM-B3LYP, N12-SX, MN12-SX. The
straight line is results for PBE global hybrids including between 0 and 70\%
exact exchange.  

The PBE global hybrid results clearly highlights the zero-sum tradeoff between
valence and core properties.  Introducing a {\em{fixed}} fraction of the
{\em{entire}} nonlocal exchange interaction increases the predicted core IP,
but also makes the predicted atomization energy less positive.  PBE0 is
near-optimal for the atomization energy, but underbinds the core electron.
PBE70 is near-optimal for the core electron, but underestimates the chemical
bond strengths.  Most global hybrids lie close to this ``zero-sum'' line,
accurately reproducing the atomization energy while underestimating the core
IP.  Long-range corrected hybrids tend to {\em{further}} underestimate the core
IP.  Screened hybrids MN12SX and N12SX, and the highly parameterized SOGGA11X,
M06-2X, and M06-HF better approach the reference value.  For this system, the
core-projected hybrid PBE0c70 clearly provides the best agreement with the
reference values employed. 

The rest of this work presents a more detailed assessment of PBE0c70 on two
data sets of core-electron ionization potentials. The first set is 33 $K$-shell
ionization potentials of second-row atoms, from 14 small molecules, referenced
to experiment.\cite{Tu2007} (In this dataset, ``MBO'' denotes 2-mercaptobenzoxazole.)  The second set is 15 $K$-shell ionization
potentials of third-row atoms, from 15 small molecules, referenced to
nonrelativistic $\Delta$MP2 calculations.\cite{Besley2021}.  Calculations use
the 6-311++G(2d,2p) basis set and B3LYP/6-311++G(2d,2p) geometries. 

Table \ref{tab:val} reports a validation of the valence properties predicted by
core-projected PBE0c70. As the goal is to recover the underlying PBE0 global
hybrid, mean absolute deviations MAD are referenced to PBE0.  Gratifyingly,
PBE0c70 gives ionization potentials within 0.01 eV of PBE0, and atomization
energies within 1 kcal/mol of PBE0.  Much larger deviations in valence
properties are seen for the PBE70 global hybrid. 

\begin{table}
\caption{\label{tab:val} Valence ionization potentials IP (eV) and atomization energies AE (kcal/mol), referenced to PBE0. }
\begin{tabular*}{\textwidth}{@{\extracolsep{\fill}}l rrr   rrr } 
\hline\hline
 & \multicolumn{3}{c}{IP} & \multicolumn{3}{c}{AE} \\ 
Molecule & PBE0 & PBE0c70 & PBE70 & PBE0 & PBE0c70 & PBE70 \\ 
\hline
CO & 11.21 & 11.21 & 14.21 & 251.6 & 251.4 & 231.4\\
H2O & 8.93 & 8.93 & 12.64 & 224.9 & 224.9 & 213.9\\
CH4 & 11.13 & 11.13 & 14.01 & 413.6 & 413.3 & 412.4\\
CH3CN & 9.75 & 9.75 & 12.23 & 294.7 & 294.0 & 268.9\\
CH3COOH & 8.16 & 8.16 & 11.53 & 792.1 & 791.4 & 761.6\\
Glycine & 7.64 & 7.64 & 10.59 & 956.1 & 955.3 & 915.6\\
MBO & 6.71 & 6.71 & 8.54 & 1734.5 & 1731.8 & 1666.5\\
PhCH3 & 7.26 & 7.26 & 8.97 & 1666.1 & 1663.8 & 1632.9\\
PhNH2 & 6.24 & 6.24 & 8.1 & 1542.7 & 1540.5 & 1500.8\\
PhOH & 6.83 & 6.83 & 8.72 & 1472.2 & 1470.1 & 1429.0\\
PhF & 7.57 & 7.57 & 9.42 & 1380.4 & 1378.4 & 1339.8\\
C2H2 & 8.73 & 8.73 & 10.96 & 399.7 & 399.0 & 384.7\\
C2H4 & 8.17 & 8.17 & 10.2 & 557.6 & 557.0 & 548.7\\
C2H6 & 9.74 & 9.74 & 12.49 & 704.8 & 704.1 & 701.8\\
AlH3 & 8.61 & 8.61 & 11 & 197.0 & 196.2 & 203.4\\
AlH2Cl & 8.75 & 8.75 & 11.24 & 217.8 & 216.9 & 221.9\\
AlH2F & 8.89 & 8.89 & 11.43 & 259.4 & 258.5 & 257.7\\
SiH4 & 9.99 & 9.99 & 12.49 & 300.5 & 299.8 & 307.6\\
H3SiOH & 8.62 & 8.62 & 11.58 & 422.5 & 421.7 & 419.6\\
H3SiCl & 9.12 & 9.12 & 11.75 & 298.3 & 297.5 & 302.8\\
PH3 & 8.02 & 8.03 & 10.19 & 222.2 & 221.8 & 221.3\\
H3PO & 7.8 & 7.8 & 10.95 & 295.0 & 294.4 & 283.0\\
H2POOH & 8.06 & 8.06 & 11.22 & 410.7 & 410.0 & 388.7\\
CH3SH & 7.1 & 7.1 & 9.45 & 456.7 & 456.2 & 451.5\\
H2CS & 6.95 & 6.95 & 9.28 & 308.5 & 308.0 & 296.4\\
H2S & 7.8 & 7.8 & 10.17 & 169.0 & 168.8 & 166.5\\
CH3Cl & 8.64 & 8.64 & 11.36 & 385.1 & 384.7 & 380.7\\
HCOCl & 9.28 & 9.28 & 12.25 & 354.6 & 354.2 & 330.5\\
HCl & 9.69 & 9.69 & 12.45 & 99.5 & 99.4 & 97.7\\
MAD &  --- & 0.00 & 2.55 &  ---- & 0.87 & 17.07 \\ 
\hline\hline
\end{tabular*}
\end{table} 

Tables \ref{tab:c1}-\ref{tab:c2} report second- and third-row core ionization
potentials for the benchmark data sets. As in previous work, PBE0 core orbital
energies are not an accurate predictor for core ionization potentials, giving 
MAD $>10$ eV for second-row atoms and $>40$ eV for 
third-row atoms.  PBE70 significantly improves the core IP. Gratifyingly,
PBE0c70 is nearly as accurate as PBE70, while maintaining PBE0 performance
for valence electron properties. 

\begin{table}
\caption{\label{tab:c1} Core ionization potentials for second-row atoms (eV).} 
\begin{tabular*}{\textwidth}{@{\extracolsep{\fill}}ll rrrr} 
\hline \hline
Molecule & Atom & Reference & PBE0 & PBE0c70 & PBE70 \\
\hline
CO & O & 542.1 & 525.9 & 549.6 & 548.4\\
& C & 295.5 & 282.7 & 300.4 & 299.1\\
H$_2$O & O & 539.9 & 523.2 & 546.9 & 545.6\\
CH$_4$ & C & 290.8 & 278.5 & 296.2 & 294.9\\
CH$_3$CN & N  & 405.6 & 392.2 & 413.0 & 411.7\\
 & {\bf{C}}N& 293.0 & 280.7 & 298.4 & 297.1\\
 & {\bf{C}}H$_3$ & 292.4 & 280.5 & 298.1 & 296.9\\
CH$_3$COOH & CO{\bf{O}}H & 540.1 & 524.5 & 548.2 & 546.9\\
 & C{\bf{O}}OH & 538.4 & 522.8 & 546.5 & 545.2\\
 & {\bf{C}}OOH & 295.4 & 283.6 & 301.3 & 300.0\\
 & CH$_3$& 291.6 & 279.7 & 297.3 & 296.1\\
Glycine & CO{\bf{O}}H & 540.2 & 524.6 & 548.4 & 547.1\\
 & C{\bf{O}}OH & 538.4 & 522.9 & 546.7 & 545.4\\
 & N & 405.4 & 391.9 & 412.6 & 411.2\\
 & {\bf{C}}OOH & 295.3 & 283.6 & 301.3 & 300.0\\
 & CH & 295.2 & 280.6 & 298.3 & 297.0\\
MBO & O & 540.6 & 525.6 & 549.3 & 548.0\\
 & N & 407.0 & 394.5 & 415.2 & 413.9\\
 & {\bf{C}}S & 295.7 & 284.5 & 302.2 & 300.9\\
 & {\bf{C}}O & 293.9 & 281.8 & 299.5 & 298.2\\
 & {\bf{C}}CN & 293.0 & 281.5 & 299.2 & 297.9\\
 & {\bf{C}}CO & 297.9 & 306.7 & 280.3 & 297.3\\
C$_6$H$_5$CH3 & CH$_3$ & 290.9 & 279.5 & 297.2 & 295.9\\
 & {\bf{C}}CH$_3$ & 290.1 & 279.3 & 296.9 & 295.7\\
C$_6$H$_5$NH2 & N & 405.3 & 392.2 & 412.9 & 411.5\\
 & {\bf{C}}N & 291.2 & 280.5 & 298.2 & 296.9\\
C$_6$H$_5$OH & O & 538.9 & 523.9 & 547.6 & 546.3\\
 & {\bf{C}}O & 292.0 & 281.3 & 298.9 & 297.6\\
C$_6$H$_5$F & F & 693.3 & 674.3 & 701.0 & 699.8\\
 & {\bf{C}}F & 292.9 & 281.9 & 299.6 & 298.3\\
C$_2$H$_2$ & C & 291.2 & 279.5 & 297.2 & 295.9\\
C$_2$H$_4$ & C & 290.7 & 279.2 & 296.9 & 295.6\\
C$_2$H$_6$ & C & 290.6 & 278.7 & 296.4 & 295.1\\
MAE & & ----& 13.0 & 7.0 & 5.3 \\ 
\hline\hline
\end{tabular*}
\end{table}

\begin{table}
\caption{\label{tab:c2} Core ionization potentials for third-row atoms (eV).}
\begin{tabular*}{\textwidth}{@{\extracolsep{\fill}}l rrrr} 
\hline \hline
Molecule & Reference & PBE0 & PBE0c70 & PBE70 \\
\hline 
AlH$_3$ & 1565.1 & 1528.5 & 1565.4 & 1566.8\\
AlH$_2$Cl & 1565.8 & 1530.0 & 1567.0 & 1568.3\\
AlH$_2$F & 1566.0 & 1529.4 & 1566.4 & 1567.7\\
SiH$_4$ & 1843.2 & 1803.3 & 1843.2 & 1844.9\\
H$_3$SiOH & 1844.0 & 1804.2 & 1844.1 & 1845.7\\
H$_3$SiCl & 1844.3 & 1805.1 & 1845.1 & 1846.7\\
PH$_3$ & 2145.8 & 2101.7 & 2144.6 & 2146.5\\
H$_3$PO & 2148.3 & 2104.7 & 2147.7 & 2149.6\\
H$_2$POOH & 2149.1 & 2105.7 & 2148.7 & 2150.6\\
CH$_3$SH & 2471.0 & 2422.8 & 2468.8 & 2471.0\\
H$_2$CS & 2471.2 & 2422.9 & 2468.9 & 2471.0\\
H$_2$S & 2471.7 & 2423.2 & 2469.2 & 2471.4\\
CH$_3$Cl & 2820.3 & 2767.4 & 2816.4 & 2818.9\\
HCOCl & 2820.6 & 2768.4 & 2817.4 & 2819.8\\
HCl & 2821.4 & 2768.2 & 2817.1 & 2819.6\\
MAE & --- & 44.1 & 1.6 & 1.3 \\ 
\hline\hline
\end{tabular*}
\end{table}

\section{Discussion} 
\label{sec:discussion}

These results motivate further exploration of projected hybrids. Core-projected
hybrids appear to be a promising choice for beyond-zero-sum TD-DFT
simulations of X-ray absorbance and fluorescence of second- and third-row
atoms, including vibronic structure, without the need for $\Delta$SCF
corrections.\cite{Fouda2020}
Other projections, for example projections onto metal $d$-electron states
within a single unit cell, could provide connections between screened hybrid
and DFT+U simulations of periodic systems. 
Going beyond the single-determinant approximation in eq \ref{eq:PH} could
provide an interesting alternative to active space selection and orbital
localization in multiconfigurational
methods.\cite{Janesko2022,Bao2019,Manni2014}  Consider for example a
calculation on a large organometallic complex known to possess multireference
character in the metal $d$ electrons. Rather than choosing an active space of
correlated MOs, one could project the metal atom $d$ AOs into the reference
system, leaving only $\sim 10^3$ nonzero AO-basis two-electron integrals.
Algorithms that account for this extreme sparsity of AO-basis integrals could
potentially provide near-full-CI accuracy for the entire reference system,
giving a ``black-box'' alternative to multireference wavefunction-in-DFT
approaches.\cite{Sharma2021} Overall, the present results motivate further
development of Adiabatic Projection hybrids, just as Refs \citenum{Heyd2003}
and \citenum{Dreuw2003} motivated broad adoption of screened and LC hybrids. 

\section{Acknowledgments} 

The author acknowledges the Texas Advanced Computing Center at the University
of Texas at Austin for providing HPC resources that have contributed to the
research results reported within this paper. 

\bibliographystyle{aip}

\end{document}